\documentclass[pdflatex,sn-mathphys-num]{sn-jnl}

\usepackage{graphicx}%
\usepackage{multirow}%
\usepackage{amsmath,amssymb,amsfonts}%
\usepackage{amsthm}%
\usepackage{mathrsfs}%
\usepackage[title]{appendix}%
\usepackage{xcolor}%
\usepackage{textcomp}%
\usepackage{manyfoot}%
\usepackage{booktabs}%
\usepackage{algorithm}%
\usepackage{algorithmicx}%
\usepackage{algpseudocode}%
\usepackage{listings}%

\newcommand{\ion}[2]{#1\,{\footnotesize\textsc{#2}}}
\newcommand{\celltwolines}[2]{\begin{tabular}[c]{@{}c@{}}$#1$\\\scriptsize$[#2]$\end{tabular}}

\theoremstyle{thmstyleone}%
\theoremstyle{thmstyletwo}%

\theoremstyle{thmstylethree}%

\begin{document}



\title[Article Title]{Nonthermal line broadening at solar flare footpoints is primarily field-aligned}
%




%



\author*[1]{\fnm{Andy S. H.} \sur{To}}\email{andysh.to@esa.int}\email{andyshto.work@gmail.com}

\author[2]{\fnm{Alexander J. B.} \sur{Russell}}\email{ar51@st-andrews.ac.uk}


\affil*[1]{\orgname{European Space Agency (ESA), European Space Research and Technology centre (ESTEC)}, \orgaddress{\street{Keplerlaan 1}, \city{Noordwijk}, \postcode{2201 AZ}, \country{The Netherlands}}}

\affil[2]{\orgdiv{School of Mathematics \& Statistics}, \orgname{University of St Andrews}, \orgaddress{\city{St Andrews}, \postcode{KY16 9SS}, \country{United Kingdom}}}



\abstract{


Magnetic reconnection powers solar and stellar flares, but a full understanding of how the released energy is transported and converted within the solar atmosphere remains elusive. One clue lies at solar-flare footpoints, where spectral lines are far broader than the electron temperature alone can explain. Unresolved flows, waves, turbulence and ion heating have all been proposed, but observations have not yet conclusively distinguished between these mechanisms. Here we perform an unprecedented geometric test for flare footpoints, using 4,593 Hinode/EIS spectra from 407 C-- to M--class flares. Line widths decrease systematically from disk centre to limb in all coronal emission lines, showing that the dominant broadening component is magnetic field aligned rather than isotropic or transverse. Cooler lines retain substantial broadening into the early decay phase, consistent with persistent unresolved field-aligned flows or line-of-sight velocity gradients. Hotter lines show an impulsive component that decays rapidly after the soft X-ray peak, consistent with preferential ion heating and ion temperature anisotropy. These findings resolve the long-standing question of the nature of line broadening at flare footpoints, place direct limits on flare energetics, and motivate a new direction in flare physics incorporating distinct field-aligned and perpendicular ion temperatures that exceed the electron temperature.
}

\keywords{Sun: flares -- Line: profiles -- Sun: corona -- Sun: chromosphere -- Turbulence -- Waves  -- Magnetic reconnection}



\maketitle

\section{Introduction}\label{sec1}

Magnetic reconnection is a fundamental energy release process across astrophysical plasmas, yet identifying how the released energy is partitioned between different physical processes in natural systems has proved challenging~\cite{Yamada2022mrms.book.....Y, Drake2025SSRv..221...27D,Sironi2025ARA&A..63..127S}. Solar flares provide one of the Solar System's most dramatic displays of reconnection, releasing up to $10^{25}$~J of magnetic energy in a few minutes, and depositing most of it at the compact dense lower atmospheric footpoints~\cite{Fletcher2011SSRv..159...19F}. A key signature of this deposition is that spectral lines formed at flare footpoints are substantially broader than the electron temperature alone can account for~\citep{Grineva1973,Doschek1979,Culhane1981}. The physical origin of this excess broadening encodes how flare energy is transported and dissipated, and whether ions are energised differently from electrons, with implications for magnetic reconnection in stellar flares and other astrophysical environments. Yet despite decades of observation, the dominant contribution to flare-footpoint line broadening has remained unknown.

Four broad classes of mechanism have been proposed, and each predicts a distinct dependence on the angle between the magnetic field and the line of sight. Unresolved field-aligned flows, in particular chromospheric evaporation, produce maximum broadening when viewing down the field (disk centre) and minimum across it (limb), whether the superposition occurs between unresolved loops~\citep{Antonucci1982SoPh...78..107A,Antonucci1984ApJ...287..917A} or along steep line-of-sight velocity gradients~\citep{Mandage2020ApJ...891..122M,Cho2024ApJ...975...33C}. However, this picture struggles to explain the appearance of broadening in hot lines before high-speed upflows~\citep{Antonucci1986ApJ...301..975A,Alexander1998} or the symmetry of fully-blueshifted footpoint profiles~\citep{Polito2019ApJ...879L..17P}. Alfv\'{e}n waves, which can transport substantial Poynting flux from the corona to the chromosphere in flares~\citep{Fletcher:2008,Russell2013ApJ...765...81R,Reep2016ApJ...818L..20R,Kerr2016,Russell2024,Lorincik2025}, carry transverse motions and therefore predict the opposite signature: the broadest line at the limb, narrowest at disk centre. Magnetohydrodynamic (MHD) turbulence, which is likely important for flare particle acceleration~\citep{Larosa1993ApJ...418..912L,Tsuneta1995,Petrosian2012SSRv..173..535P,Klein2017SSRv..212.1107K} and has been identified spectroscopically at both looptops~\citep{Kontar2017PhRvL.118o5101K} and footpoints~\citep{Jeffrey2018SciA....4.2794J,Chitta2020}, could produce the same centre-to-limb variation if dominated by transverse motions, but no viewing angle dependence if isotropic. Finally, preferential ion heating during reconnection, recently argued by \citet{Russell2025ApJ...990L..39R} to raise $T_i$ to several times $T_e$ in parts of hot flare plasma, and previously suggested for the non-flaring corona~\citep{Seely1997,Tu1998,Landi2007}, produces a viewing-angle dependence only where the ion temperature itself is anisotropic: broader at disk centre when $T_{i\parallel}>T_{i\perp}$ and broader at limb when $T_{i\perp}>T_{i\parallel}$.

The viewing-angle dependence therefore provides the most direct empirical test of which line-broadening mechanism dominates. The largest centre-to-limb surveys of flare line widths to date, performed using soft X-ray spectra from Yohkoh's Bragg Crystal Spectrometer (BCS), found no significant trend for the \ion{Fe}{XXV}, \ion{Ca}{XIX}, and \ion{S}{XV} resonance lines during the impulsive to early gradual phase~\citep{Mariska1993ApJ...419..418M,Mariska1994ApJ...434..756M}. Those observations, however, averaged over the entire flaring loop system with widely varying magnetic field angles to the line of sight within individual flares. Any centre-to-limb signature of the broadening at footpoints is therefore likely to have been washed out.

\begin{figure*}
    \centering
    \includegraphics[width=\linewidth]{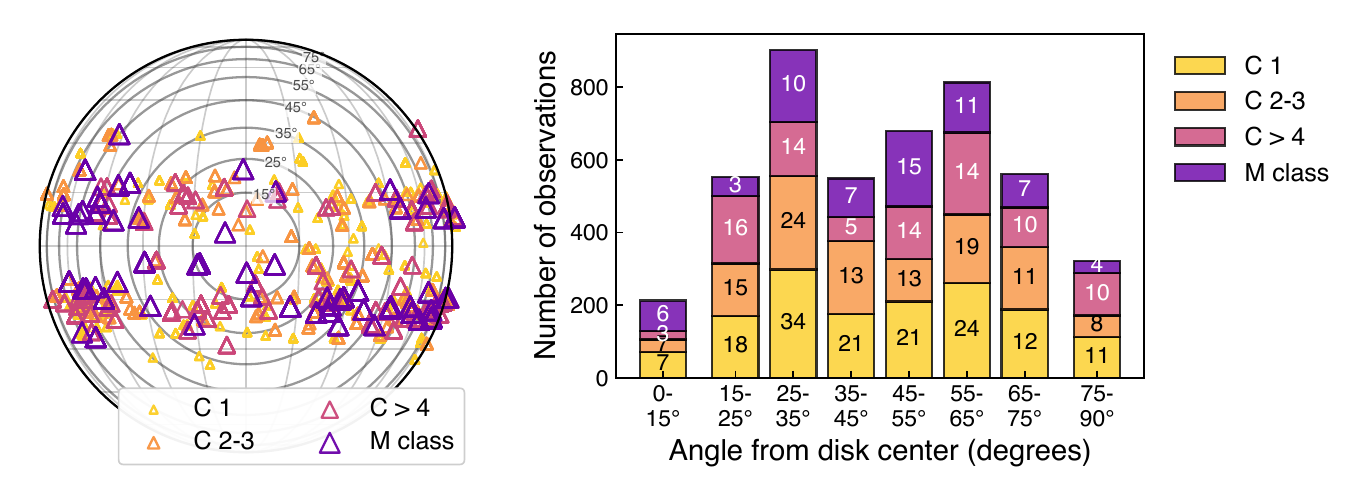}
    \caption{Distribution of EIS flare observations versus viewing angle from disk centre for C1 (yellow), C2--3 (orange), C$>$4 (pink), and M class (purple) flares within $\pm$5 minutes of the GOES soft X-ray peak. (left) Heliographic positions. (right) Histogram of number of individual EIS observations (4,593 in total). Numbers indicate the number of distinct flares per viewing-angle bin (407 in total).}
    \label{fig:location_angle}
\end{figure*}

The Hinode/Extreme-ultraviolet Imaging Spectrometer (EIS; \citealt{Culhane2007Jun}) removes this limitation by spatially isolating footpoint emission across a broad temperature range, from \ion{Fe}{X} (0.9~MK) to \ion{Fe}{XXIV} (17.8~MK). Combined with the comprehensive flare catalogue compiled by \citet{To2025ApJ...993..102T}, this enables the first centre-to-limb survey of flare footpoint line broadening across a broad electron temperature range from \ion{Fe}{X} to \ion{Fe}{XXIV}.

Here we exploit 4,593 footpoint spectra from 407 C- to M-class flares, distributed from disk centre to near the limb (Fig.~\ref{fig:location_angle}), to determine how excess broadening depends on viewing angle. We find a systematic decrease in line width from disk centre to limb across all Fe ions analysed, showing that the dominant component of excess broadening is field-aligned. This result disfavours isotropic turbulence and transverse mechanisms such as Alfv\'en waves as the primary origin of the excess widths. Cooler lines retain most of their broadening into the early decay phase, consistent with persistent unresolved field-aligned flows or line-of-sight velocity gradients. Hotter lines show an additional impulsive enhancement that decays rapidly after the soft X-ray peak, suggesting an additional impulsive field-aligned component that may reflect anisotropic ion heating with $T_{i\parallel}>T_{i\perp}>T_e$.

\section{Results}\label{sec2}

\begin{figure*}
    \centering
    \includegraphics[width=0.95\linewidth]{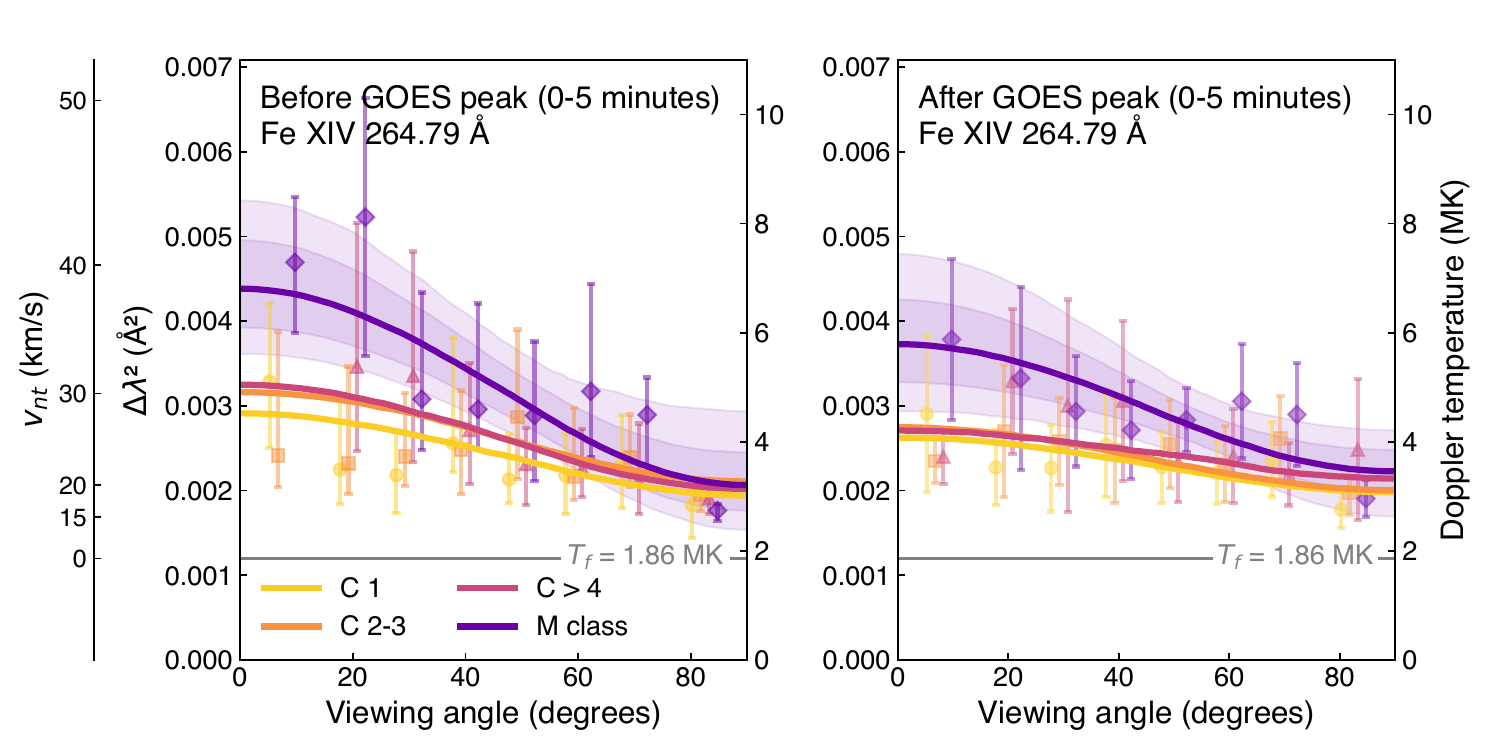}
    \caption{Centre-to-limb variation of the \ion{Fe}{XIV}~264.79~\AA\ line width during the 0--5~minute intervals before (left) and after (right) the GOES soft X-ray peak. Markers show median values and error bars indicate the 25th--75th percentile range. Solid curves show the best-fit bi-Maxwellian model for each flare class; for M-class flares, the dark and light shaded regions indicate the 68\% and 95\% confidence intervals, respectively. The horizontal grey line marks the equilibrium formation temperature of the line ($T_f = 1.86$~MK). The three vertical axes show the line width squared, $\Delta\lambda^2$, the Doppler temperature, $T_D$, and the corresponding nonthermal velocity, $v_\mathrm{nt}$.}
    \label{fig:fe_14_comparison2}
\end{figure*}

\begin{figure*}
    \centering
    \includegraphics[width=0.95\linewidth]{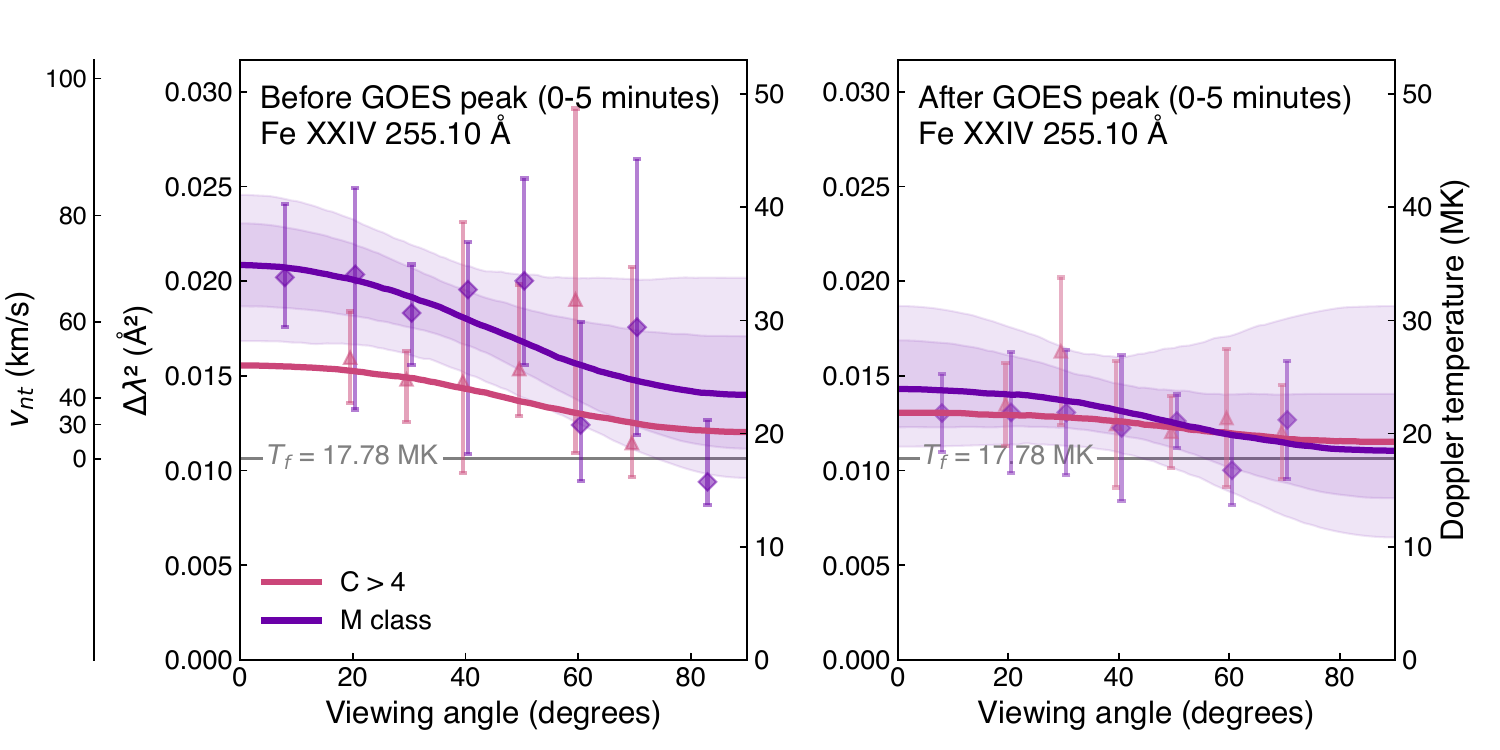}
    \caption{Centre-to-limb variation of the \ion{Fe}{XXIV}~255.10~\AA\ line width during the 0--5~minute intervals before (left) and after (right) the GOES soft X-ray peak. Symbols, curves, and shaded regions are defined as in Fig.~\ref{fig:fe_14_comparison2}. The horizontal grey line marks the equilibrium formation temperature of the line ($T_f = 17.78$~MK). Only C$>$4 and M-class flares are shown because of the low signal-to-noise of \ion{Fe}{XXIV} emission in smaller events.}
    \label{fig:fe_24_comparison}
\end{figure*}

\subsection{Line widths decrease from disk centre to limb}

We first examine whether flare-footpoint line widths vary systematically with viewing angle across the solar disk. Our analysis comprises 4,593 EIS footpoint observations from 407 C- to M-class flares spanning 2011--2024, taken within $\pm 5$~minutes of the GOES soft X-ray peak and distributed across the solar disk (Fig.~\ref{fig:location_angle}), drawn from the flare catalogue of \citet{To2025ApJ...993..102T} (Methods).

Figure~\ref{fig:fe_14_comparison2} shows a clear centre-to-limb decrease for the \ion{Fe}{XIV}~264.79~\AA\ line during the $\pm 5$~minute windows around the GOES soft X-ray peak. Across all flare classes, the broadening is largest at disk centre and decreases systematically toward the limb. The trend is strongest before the GOES peak in M-class flares, where the disk-centre broadening of $\Delta\lambda^2 = 0.0044$~\AA$^2$ ($T_D = 6.8$~MK, $v_\mathrm{nt} = 38$~km~s$^{-1}$) falls to $\Delta\lambda^2 = 0.0022$~\AA$^2$ ($T_D = 3.3$~MK, $v_\mathrm{nt} = 21$~km~s$^{-1}$) at the limb, a factor-of-two reduction in $\Delta\lambda^2$ across the disk. After the GOES peak, the same centre-to-limb decrease persists but with lower disk-centre values of $\Delta\lambda^2 = 0.0037$~\AA$^2$ decreasing to $\Delta\lambda^2 = 0.0022$~\AA$^2$ at the limb.

\subsection{Anisotropic broadening spans ionisation states}\label{sec:temperature_span}

The same centre-to-limb trend is present in every Fe line examined, covering nearly 1.5 orders of magnitude in equilibrium formation temperature, $T_f$. Figure~\ref{fig:fe_24_comparison} shows the equivalent behaviour for the hot \ion{Fe}{XXIV}~255.10~\AA\ line (formation temperature 17.78~MK), which is the hottest line in our sample. For M-class flares, the fitted \ion{Fe}{XXIV} nonthermal velocity decreases from $v_{\rm nt}\simeq72$~km~s$^{-1}$ at disk centre to $v_{\rm nt}\simeq40$~km~s$^{-1}$ near the limb before the GOES peak, and from $v_{\rm nt}\simeq38$~km~s$^{-1}$ to $v_{\rm nt}\simeq13$~km~s$^{-1}$ after the peak. Only C$>$4 and M-class flares are shown because \ion{Fe}{XXIV} is too weak for reliable measurements in smaller events. The same behaviour is seen in all other lines analysed, including \ion{Fe}{X}, \ion{Fe}{XII}, \ion{Fe}{XVI}, and \ion{Fe}{XXIII} (Extended Data Fig.~\ref{fig:M_class_all_lines_pos} and Fig.~\ref{fig:M_class_all_lines_post}). A systematic centre-to-limb decrease in line width is therefore a general feature of flare footpoint emission for Fe ions with formation temperatures from 0.9~MK to 17.8~MK.

\subsection{Quantifying the dominant field-aligned component}

To quantify this anisotropy, we model the viewing-angle dependence of the observed line widths with a bi-Maxwellian population of emitting ions. For a line of sight observation making an angle $\theta$ to the magnetic field, the effective Doppler temperature is
\begin{equation}\label{equ:TDTP_fit}
T_D = \frac{T_{\perp}}{1 + \mu\cos^2\theta - \mu^2\sin^2\theta\cos^2\theta/(1 + \mu\sin^2\theta)},
\end{equation}
where
\begin{equation}
\mu = \frac{T_{\perp}}{T_{\|}} - 1.
\end{equation}
$\mu = 0$ corresponds to isotropic broadening, $\mu > 0$ when $T_{\perp} > T_{\parallel}$, and $\mu < 0$ when $T_{\parallel} > T_{\perp}$. The same definition of $\mu$ appears in the stability criteria for mirror and firehose instabilities and is often used in solar wind studies of temperature anisotropy, e.g., \citet{Matteini2006JGRA..11110101M}. Equation~(\ref{equ:TDTP_fit}) applies whether the anisotropy reflects a true ion-temperature anisotropy, unresolved plasma motions with a bi-Maxwellian distribution, or a combination of the two. It was fitted to the data using a stratified hierarchical bootstrap that accounts for flare-to-flare variability and uneven sampling across viewing angles and flare magnitudes (Appendix~\ref{app:bootstrap_method}).

Table~\ref{tab:M_class_anisotropy_fit} reports the best-fit $\mu$, $T_\perp$ and $T_\parallel$ for M-class flares in six Fe lines from 0.9 to 17.8~MK, before and after the GOES peak, with 95\% confidence intervals from the bootstrap. The inferred $\mu$ is negative in every line and in both phases, showing that the dominant broadening component is field-aligned ($T_\parallel > T_\perp$) rather than isotropic or perpendicular. Extended Data Tables~\ref{tab:anisotropy_fit_all_before} and~\ref{tab:anisotropy_fit_all_after} report the equivalent fits for all flare classes.

\subsection{Anisotropy is persistent for cool lines and impulsive for hot lines}\label{sec:cool_vs_hot}

The strength and persistence of the anisotropy differ between the warm coronal lines and the hottest flare lines. For \ion{Fe}{X}, \ion{Fe}{XII}, \ion{Fe}{XIV} and \ion{Fe}{XVI}, which have formation temperatures from 0.9 to 2.6~MK, $\mu$ is significantly negative at the 95\% confidence level both before and after the GOES peak (Table~\ref{tab:M_class_anisotropy_fit}). The field-aligned excess line width in these lines therefore persists into the early decay phase (as seen for \ion{Fe}{XIV} in Fig.~\ref{fig:fe_14_comparison2}).

The hotter lines behave differently. For \ion{Fe}{XXIII} and \ion{Fe}{XXIV}, formed at 14.1~MK and 17.8~MK, the field-aligned anisotropy is significant before the GOES peak. After the peak, the fits remain consistent with field-aligned anisotropy but are no longer statistically conclusive. The field-aligned Doppler temperature $T_\parallel$ also drops substantially between the pre- and post-peak windows for these hot lines, consistent with the well-known rapid decay of nonthermal widths in hot flare lines after the soft X-ray peak \citep{Doschek1990,Antonucci1999,Russell2024,Russell2025ApJ...990L..39R}. The cool and hot lines therefore behave differently in time: the field-aligned broadening in the cool lines is present across both phases, whereas the field-aligned broadening in the hot lines is predominantly impulsive.

\begin{table*}
\caption{Anisotropic broadening parameters derived from the bi-Maxwellian model fit (M class). Values are quoted as bootstrap medians with 95\% confidence intervals. The $p$ values are obtained from a separate residual-bootstrap test of the isotropic null hypothesis (Extended Data B.1)}
\label{tab:M_class_anisotropy_fit}
\centering

\scriptsize
\renewcommand{\arraystretch}{1.15}
\setlength{\tabcolsep}{4pt}
\begin{tabular}{l c c c c c c}
\toprule
\multicolumn{7}{c}{\textbf{Before GOES peak} ($-5$ to $0$ min)} \\
\midrule
\textbf{Ion} & $T_f$ (MK) & $N$ & $\mu$ & $p$ & $T_{\perp}$ (MK) & $T_{\parallel}$ (MK) \\
\midrule

\ion{Fe}{X} 184.54 & 0.9 &
363 &
\celltwolines{-0.425}{-0.914,\,-0.215} &
0.027 &
\celltwolines{4.53}{0.80,\,6.56} &
\celltwolines{7.89}{5.54,\,13.97} \\

\ion{Fe}{XII} 192.39 & 1.3 &
398 &
\celltwolines{-0.427}{-0.721,\,-0.076} &
0.001 &
\celltwolines{2.99}{1.64,\,4.15} &
\celltwolines{5.22}{3.43,\,6.79} \\

\ion{Fe}{XIV} 264.79 & 1.9 &
420 &
\celltwolines{-0.393}{-0.682,\,-0.278} &
0.001 &
\celltwolines{3.92}{2.39,\,4.56} &
\celltwolines{6.46}{5.61,\,8.43} \\

\ion{Fe}{XVI} 262.98 & 2.6 &
409 &
\celltwolines{-0.327}{-0.534,\,-0.187} &
0.001 &
\celltwolines{5.59}{4.30,\,6.53} &
\celltwolines{8.30}{7.21,\,10.12} \\

\ion{Fe}{XXIII} 263.77 & 14.1 &
188 &
\celltwolines{-0.378}{-0.527,\,0.043} &
0.002 &
\celltwolines{22.65}{17.44,\,31.62} &
\celltwolines{36.40}{27.83,\,40.50} \\

\ion{Fe}{XXIV} 255.10 & 17.8 &
178 &
\celltwolines{-0.273}{-0.572,\,0.088} &
0.012 &
\celltwolines{25.65}{16.08,\,33.76} &
\celltwolines{35.28}{28.17,\,41.09} \\

\bottomrule
\end{tabular}

\vspace{0.8em}

\begin{tabular}{l c c c c c c}
\toprule
\multicolumn{7}{c}{\textbf{After GOES peak} ($0$ to $+5$ min)} \\
\midrule
\textbf{Ion} & $T_f$ (MK) & $N$ & $\mu$ & $p$ & $T_{\perp}$ (MK) & $T_{\parallel}$ (MK) \\
\midrule

\ion{Fe}{X} 184.54 & 0.9 &
332 &
\celltwolines{-0.343}{-0.942,\,-0.154} &
0.034 &
\celltwolines{3.69}{0.63,\,4.22} &
\celltwolines{5.61}{3.70,\,12.13} \\

\ion{Fe}{XII} 192.39 & 1.3 &
413 &
\celltwolines{-0.340}{-0.777,\,-0.095} &
0.016 &
\celltwolines{2.99}{1.47,\,3.56} &
\celltwolines{4.53}{3.41,\,7.45} \\

\ion{Fe}{XIV} 264.79 & 1.9 &
399 &
\celltwolines{-0.197}{-0.621,\,-0.175} &
0.074 &
\celltwolines{4.10}{2.63,\,4.21} &
\celltwolines{5.11}{4.56,\,7.44} \\

\ion{Fe}{XVI} 262.98 & 2.6 &
390 &
\celltwolines{-0.143}{-0.454,\,-0.013} &
0.020 &
\celltwolines{5.76}{4.37,\,6.27} &
\celltwolines{6.72}{5.91,\,8.38} \\

\ion{Fe}{XXIII} 263.77 & 14.1 &
208 &
\celltwolines{-0.151}{-0.427,\,0.442} &
0.003 &
\celltwolines{18.99}{15.06,\,28.71} &
\celltwolines{22.36}{18.06,\,27.41} \\

\ion{Fe}{XXIV} 255.10 & 17.8 &
199 &
\celltwolines{-0.190}{-0.639,\,0.516} &
0.002 &
\celltwolines{18.38}{10.82,\,31.26} &
\celltwolines{22.70}{18.85,\,31.27} \\

\bottomrule
\end{tabular}
\end{table*}

\section{Discussion}

Our large scale analysis shows that the dominant component of flare-footpoint excess line broadening is field-aligned from \ion{Fe}{X} (0.9~MK) to \ion{Fe}{XXIV} (17.8~MK). The observed broadening exhibits a clear centre-to-limb variation in line widths at the resolution of EIS. Mechanisms that produce predominantly transverse broadening, such as Alfv\'{e}n wave turbulence, cannot explain the observed anisotropy because they predict the opposite centre-to-limb behaviour. Isotropic MHD turbulence and a purely isotropic $T_i > T_e$ similarly cannot account for it by themselves, because both predict no viewing-angle dependence. The absence of a clear detection of this trend in earlier Yohkoh/BCS surveys~\citep{Mariska1993ApJ...419..418M,Mariska1994ApJ...434..756M} is consistent with BCS having averaged over entire flaring loops.

The simplest physical interpretation of the cool-line signal is a superposition of unresolved field-aligned plasma motions, most plausibly chromospheric evaporation. EIS's 3 arcsecond resolution\footnote{Ugarte-Urra 2016 -- EIS Software Note no. 8, \url{https://solarb.mssl.ucl.ac.uk/SolarB/eis_docs/eis_notes/08_COMA/eis_swnote_08.pdf}} is too coarse to resolve the fundamental scales of evaporation, and the $v_\mathrm{nt} = 38$~km~s$^{-1}$ seen at disk centre in M-class \ion{Fe}{XIV} is consistent with expected upflow speeds. This classic picture has been challenged, however, by \citet{Polito2019ApJ...879L..17P}, who showed that multi-stranded evaporation models predict asymmetric profiles with both blueshifted upflows and redshifted downflows, whereas the flare-footpoint \ion{Fe}{XXI}~1354.1~\AA\ profiles observed by IRIS are often symmetric and fully blueshifted \citep{2015ApJ...807L..22G,2016ApJ...816...89P}. Field-aligned velocity gradients along the line of sight offer a reconciling alternative: hydrodynamic simulations by \citet{Mandage2020ApJ...891..122M,Mandage2021ApJ...908..128M} show that such gradients can produce both symmetric and asymmetric hot-line profiles within the first minute of a flare. Our data do not distinguish between unresolved upflows and line-of-sight velocity gradients, but jointly, these hypotheses are sufficient to account for the cool-line excess. The persistence of these widths into the early decay phase further indicates that the field-aligned contribution continues beyond the impulsive energy input, consistent with continuation of chromospheric evaporation into the decay phase. At cool electron temperatures and typical footpoint densities~\citep{Milligan2011ApJ...740...70M}, ion and electron temperatures equilibrate in fractions of a second (Methods), so a significant $T_i > T_e$ contribution seems unlikely for these lines.

The hot \ion{Fe}{XXIII} and \ion{Fe}{XXIV} lines require an additional impulsive-phase contribution. Their field-aligned Doppler temperatures drop sharply between the pre- and post-peak windows, approaching cool Fe line values after the GOES peak. At $T_e \sim 18$~MK and $n_e \sim 10^{16}$~m$^{-3}$, proton--electron thermal equilibration takes $\sim100$~s, as recently highlighted by \citet{Russell2025ApJ...990L..39R}, and proton temperature anisotropies relax on $\sim 15$--$20$~s (Methods); both are long enough for distinct parallel and perpendicular ion temperatures to develop. Taken with the observed $T_\parallel > T_\perp$, this points to preferential parallel ion heating with $T_{i\parallel} > T_{i\perp} > T_e$ in the plasma where \ion{Fe}{XXIII} and \ion{Fe}{XXIV} are formed. 
Ion cyclotron heating~\citep{Bahauddin2021} predicts $T_{i\perp} > T_{i\parallel}$ and is therefore disfavoured as the dominant heating process within the flare footpoints. In contrast, in situ observations, PIC simulations and theoretical analyses of collisionless reconnection have found that parallel ion heating exceeds perpendicular ion heating~\citep{Drake2009,2014GeoRL..41.7002P,2015GeoRL..42.7239H}. The inferred $T_\parallel/T_\perp$ ratios for \ion{Fe}{XXIII} and \ion{Fe}{XXIV}, of 1.6 and 1.4 respectively, are less than the $\Delta T_\parallel/\Delta T_\perp\sim 2$ found in collisionless environments~\citep{2014GeoRL..41.7002P}, which may be ascribed to isotropisation by collisions in the  corona. Our results therefore add further support to Russell's proposition \citep{Russell2025ApJ...990L..39R} that ion heating by reconnection can account for a significant part of the excess line widths of hot flare lines. 

We also remark that temperature anisotropies can result from other flare processes. The CGL double-adiabatic relation (Methods) predicts $T_\parallel/T_\perp \propto n^2/B$ along an expanding flux tube, so that chromospheric evaporation into an expanding post-flare tube naturally develops $T_{\parallel} > T_{\perp}$ provided the plasma density falls more slowly than $B^{1/2}$. Additionally, a superposition of a field-aligned ion beam and a background population has an enhanced parallel kinetic temperature (interpreted as the second-order velocity moment, rather than a Maxwellian distribution). Therefore, while ion heating by magnetic reconnection provides a sufficient explanation for the data, future work might also investigate these additional sources of parallel heating. 

A modest residual broadening is present at the limb in every line ($\sim 20$~km~s$^{-1}$ for the cool lines), indicating that a weak non-field-aligned component coexists with the dominant field-aligned one. This may represent a modest contribution from Alfv\'en waves or MHD turbulence, and in the hot Fe lines could instead be accounted for by $T_{i\parallel}>T_{i\perp}>T_e$. Finally, we remark that EIS's spatial and spectral resolutions are lower than those of IRIS and the upcoming MUSE and Solar-C EUVST. Repeating the centre-to-limb analysis with those instruments will test whether cool-line profiles are intrinsically symmetric and fully blueshifted, as has often been seen for IRIS \ion{Fe}{XXI} \citep{2015ApJ...807L..22G,2016ApJ...816...89P,Polito2019ApJ...879L..17P}, further constraining the unresolved-flow picture.

The origin of excess line widths in solar flare footpoints has long been undetermined. Our study provides evidence that it is predominantly field-aligned, across flare classes and a broad range of ion formation temperatures. This rules out explanations in terms of unresolved transverse waves, isotropic turbulence and ion cyclotron heating. This has significant consequences for energy transport and dissipation in response to magnetic reconnection, as the Poynting flux and energy density of Alfv\'en wave turbulence are now limited by the smaller values of $v_{nt}$ obtained for the limb, reducing them by a factor greater than three. Finally, we have proposed that the rapid decrease in the \ion{Fe}{XXIII} and \ion{Fe}{XXIV} Doppler temperatures near disk centre, and their centre-to-limb variation, are consistent with non-equilibrium temperatures with $T_{i||} > T_{i\perp} > T_e$. This is consistent with Russell's proposition \citep{Russell2025ApJ...990L..39R} that ion heating by magnetic reconnection can account for a significant part of the excess line widths of hot flare lines, extending its application to flare footpoints and adding the support of agreement between predicted and observed anisotropy. While species temperature differences and ion temperature anisotropies are well documented for low-collisionality plasmas such as the solar wind, our findings motivate reconsideration of fundamental modelling assumptions in denser environments of flares.

\section{Methods}\label{sec:Method}

\subsection{Data and line fitting}


The Hinode/EIS line width dataset compiled by \citet{To2025ApJ...993..102T} contains 1,449 C to X class flares, from 2011 to 2024, with line widths measured at flare footpoints across multiple Fe lines. Emission lines from \ion{Fe}{X} (0.9~MK) to \ion{Fe}{XXIV} (17.8~MK) were fitted as described by \citet{To2025ApJ...993..102T}. The \ion{Fe}{XXIII} line was fitted with two Gaussian profiles to disentangle the contribution from an unidentified blend in the blue wing~\citep{Polito2017A&A...601A..39P,Stores2021ApJ...923...40S,DelZanna2025MNRAS.544.2513D} or possible effect of time resolution sampling~\citep{Mandage2020ApJ...891..122M, Mandage2021ApJ...908..128M}. The \ion{Fe}{XXIV}~255.10~\AA\ line was fitted with a single Gaussian. Following the quality control procedures established in \citet{To2025ApJ...993..102T}, we filter out spectral fits where the fitted intensity is $<1~\mathrm{erg~cm^{-2}~s^{-1}~sr^{-1}}$ and/or where the fitting error exceeds 10\% of the fitted intensity.

\subsection{Doppler temperatures and nonthermal velocities}



To avoid implicitly assuming $T_i=T_e$ or ionisation equilibrium, our centre-to-limb analysis results are presented using the instrument-corrected line width $\Delta\lambda$, the Doppler temperature $T_D$ and the nonthermal velocity $v_{nt}$. The $\Delta\lambda$ is obtained by subtracting the instrumental width\footnote{Young 2011 -- EIS Software Note no. 7, \url{https://solarb.mssl.ucl.ac.uk/SolarB/eis_docs/eis_notes/07_LINE_WIDTH/eis_swnote_07.pdf}} from the fitted width, in quadrature. The Doppler temperature is then defined as 
\begin{align}
    T_{D} = \frac{m_ic^2}{2k_B\lambda^2}\frac{\Delta\lambda_{\rm fit}^2 - \Delta\lambda^2_{\mathrm{inst}}}{4~\mathrm{ln}2},
\end{align}
where $m_i$ is the mass of the element, $c$ is the speed of light, $k_B$ is the Boltzmann's constant and $\lambda$ is the wavelength of the line. 
Finally, for interpretation in terms of unresolved motions,
\begin{align}
    v_{nt}=\sqrt{\frac{2 k_B(T_D - T_e)}{m_i}},
\end{align}
where $T_e$ is taken as the equilibrium formation temperature of the spectral line from the \texttt{SolarSoftWare (SSW)} \texttt{eis\_width2velocity.dat} file~\citep{Mazzotta1998A&AS..133..403M}.

\backmatter





\bmhead{Acknowledgements}

A.S.H.T. acknowledges support through the European Space Agency (ESA) Research Fellowship Programme in Space Science. A.J.B.R. acknowledges funding by the UK's Science and Technology Facilities Council (STFC) Consolidated Grant ST/W001195/1. A.S.H.T. thanks Deborah Baker for insightful conversations and Adam Finley, Antonio La Marca, and Jo Ann Egger for helpful comments on the plots. A.J.B.R. acknowledges valuable conversations with Vanessa Polito, Paola Testa and Bart De Pontieu, including as part of NASA HGI grant No. 80NSSC20K0716. Hinode is a Japanese mission developed and launched by ISAS/JAXA, collaborating with NAOJ as a domestic partner, and NASA and STFC (UK) as international partners. Scientific operation of Hinode is performed by the Hinode science team organised at ISAS/JAXA. This team mainly consists of scientists from institutes in the partner countries. Support for the post-launch operation is provided by JAXA and NAOJ (Japan), STFC (UK), NASA, ESA, and NSC (Norway). This research has made use of NASA's Astrophysics Data System Bibliographic Services. Both authors thank the organisers of the Hinode 18 conference for facilitating the start of this work.

\section*{Declarations}


\subsection*{Data availability}

The raw Hinode/EIS data analysed in this study are publicly available from the NRL Hinode/EIS archive at \url{https://eis.nrl.navy.mil/}. The flare-footpoint line width and nonthermal velocity measurements used as the basis for this work are available via Zenodo at \url{https://doi.org/10.5281/zenodo.15613861}~\cite{To2025_Zenodo_dataset}. The derived centre-to-limb analysis dataset, including, for example, the Doppler temperatures, viewing angles, flare classifications and data underlying the figures and tables in this manuscript, is available via Zenodo at \url{https://doi.org/10.5281/zenodo.20492139}~\cite{To2026_center2limb_dataset}.

\subsection*{Code availability}
The custom Python scripts used to construct the derived centre-to-limb dataset, perform the bi-Maxwellian fits, carry out the bootstrap uncertainty estimates and generate the figures are available via Zenodo at \url{https://doi.org/10.5281/zenodo.20492139}~\cite{To2026_center2limb_dataset}. For reading and preparing EIS data, we used the \texttt{EISPAC v0.97.2} Python package, available via GitHub (\url{https://github.com/USNavalResearchLaboratory/eispac}; \citealt{Weberg2023JOSS....8.4914W}). This research also used the open-source Python packages \texttt{matplotlib v3.9.2}~\cite{Hunter2007CSE.....9...90H}, \texttt{pandas v2.2.2}~\cite{mckinney-proc-scipy-2010,reback2020pandas}, \texttt{numpy v2.0.1}~\cite{Harris2020Natur.585..357H}, \texttt{scipy v1.16.3}~\cite{Virtanen2020NatMe..17..261V} and \texttt{Jupyter Notebook v7.2.1}~\cite{jupyter2007CSE.....9c..21P,kluyver2016jupyter}.

\subsection{Funding}
A.S.H.T. acknowledges support through the European Space Agency (ESA) Research Fellowship Programme in Space Science. A.J.B.R. acknowledges funding by the UK's Science and Technology Facilities Council (STFC) Consolidated Grant ST/W001195/1. 

\subsection{Competing interests}
The authors declare no competing interests.

\subsection{Author contributions}
The research was designed and conducted jointly by A.S.H.T. and A.J.B.R.
A.S.H.T. curated the EIS line widths dataset, led the statistical analysis and wrote the software.
A.J.B.R. suggested exploring centre-to-limb variation and led the interpretation of results, particularly with regard to species temperatures, temperature anisotropy and mathematical modelling.
Both authors contributed to conceptualisation, study design, creation of figures and writing of the manuscript.




\noindent

\bigskip
\begin{flushleft}%




\end{flushleft}

\begin{appendices}

\section{Extended Data: centre-to-limb line width variation across all lines}

\begin{figure*}
    \centering
    \includegraphics[width=\linewidth]{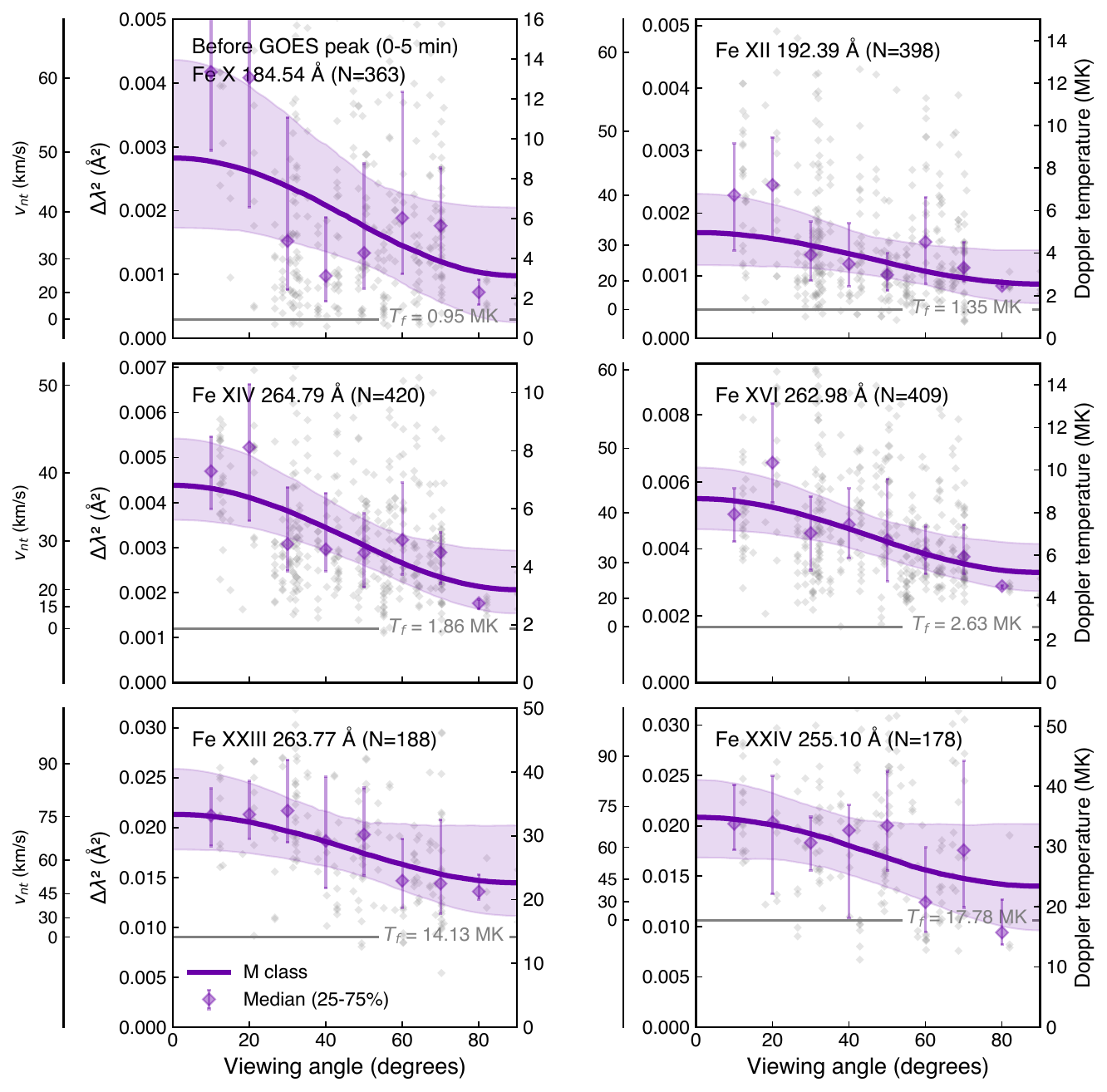}
    \caption{Centre-to-limb variation for M class flares during 0--5 minutes before GOES peak across six \ion{Fe}{} lines spanning $\log(T/K) \sim 6.0$--7.3. Purple diamond points show medians with 25th--75th percentile error bars; grey points show individual observations. Purple curves show best-fit bi-Maxwellian models with 95\% (2$\sigma$) confidence intervals (shaded area). Grey horizontal lines mark formation temperatures.}
    \label{fig:M_class_all_lines_pos}
\end{figure*}

\begin{figure*}
    \centering
    \includegraphics[width=\linewidth]{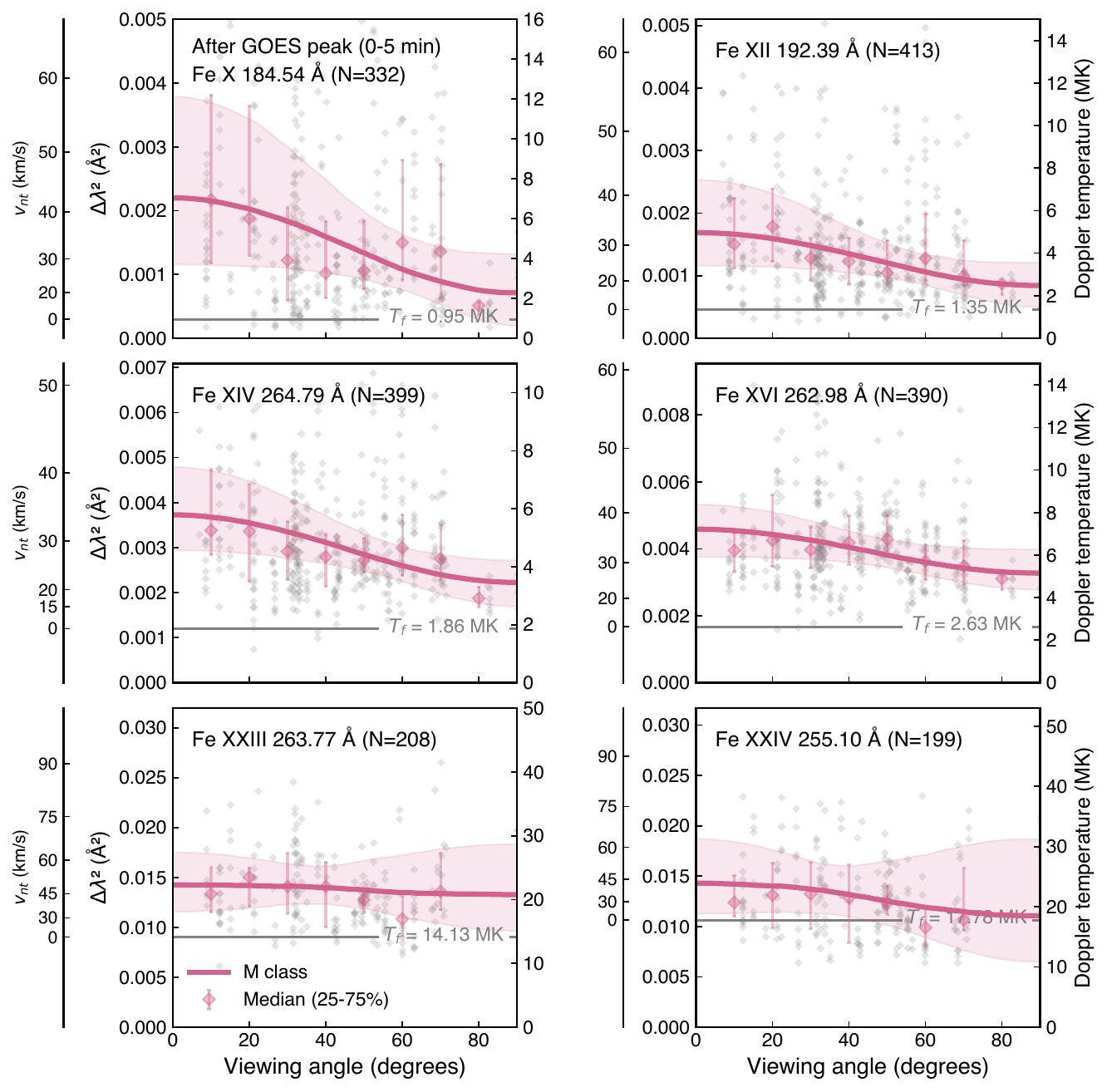}
    \caption{Same as Figure~\ref{fig:M_class_all_lines_pos} but for 0--5 minutes after GOES soft X-ray peak.}
    \label{fig:M_class_all_lines_post}
\end{figure*}

Figures~\ref{fig:M_class_all_lines_pos} and \ref{fig:M_class_all_lines_post} show the centre-to-limb variation of line broadening across six \ion{Fe}{} lines spanning $\log(T_f/K) \sim 6.0$--7.3, measured during the 0--5 minutes before and after the GOES soft X-ray peak for M class flares, respectively. Before the GOES peak, a clear and consistent centre-to-limb variation is present for all the ionisation states. The centre-to-limb variation is generally less pronounced after the peak than before, mainly due to reduction of $\Delta \lambda^2$ at disk centre between the pre- and post-peak phases. For the hot flare lines \ion{Fe}{XXIII} and \ion{Fe}{XXIV}, confidence intervals in the post-peak phase span both negative and positive $\mu$, so while their median fits have negative $\mu$ those particular data are also consistent with isotropic line broadening, unlike in the pre-peak phase.

\section{Extended Data}
\subsection{Fits, uncertainty estimation via stratified bootstrap, and hypothesis testing for anisotropy}
\label{app:bootstrap_method}

To estimate uncertainties in the fitted parameters $T_{\perp}$, $T_{\parallel}$, and $\mu$, we employ a stratified hierarchical bootstrap resampling method \citep{Efron1979bootstrap} designed to mitigate several sampling biases. The data have a clustered structure: each flare contributes a variable number of spectral measurements (from a few to several dozen), and the distribution of flares across viewing angles is mildly non-uniform (Figure~\ref{fig:location_angle}). Without adjustment, densely sampled flares and overrepresented angular ranges would dominate the fit and lead to underestimated uncertainties.

We therefore implement a two-dimensional stratified bootstrap that balances both viewing angle and flare magnitude. For each flare category we partition flares into cells defined by five viewing-angle bins spanning $0$--$90^{\circ}$ and two flare-magnitude bins. For each of 1,000 bootstrap iterations, we draw one flare with replacement from each non-empty cell to enforce balanced coverage in $(\theta,\mathrm{flare\ magnitude})$ space. Within each selected flare, we then draw a fixed number of 15 observations with replacement. This hierarchical design limits the influence of any single flare. Each bootstrap sample is refit using the same procedure as for the original data (Equation~\ref{equ:TDTP_fit}), yielding a distribution for $(T_{\perp},\mu)$ from which we report medians and 95\% confidence intervals as uncertainties.


As a complementary significance test, we ask whether an apparent non-zero $\mu$ could be produced by flare-correlated scatter even when the underlying distribution is isotropic ($\mu=0$). We first fit the model to the data with $\mu$ fixed to zero and compute the residuals about this isotropic fit. We then generate synthetic datasets using a cluster residual bootstrap. Flares are resampled with replacement, residuals are resampled within each selected flare, and these residuals are added to the best fitting null model evaluated at the corresponding observed viewing angles. Each synthetic dataset is refit with the full anisotropic model, producing a null distribution of $\hat{\mu}^{\ast}$. Repeating this procedure for 2,000 realisations gives a two-sided bootstrap p-value, computed as the fraction of successful bootstrap refits for which the absolute value of the bootstrap anisotropy, $|\hat{\mu}^{\ast}|$, is greater than or equal to the absolute value of the measured anisotropy, $|\hat{\mu}|$. This test therefore asks whether the measured anisotropy is more extreme than expected from flare-correlated residual scatter under the isotropic model.

\subsection{Thermal equilibration and temperature-anisotropy relaxation}
The possibility for $T_i>T_e$ in solar flares, proposed by \citet{Russell2025ApJ...990L..39R} (Discussion), rests on how rapidly Coulomb collisions drive the proton and electron populations toward thermal and isotropic equilibrium at flare densities and temperatures. The relevant rates are taken from the NRL Plasma Formulary~\citep{NRL} in the limit of small temperature differences (the null hypothesis).
The proton--electron thermal equilibration rate scales as
\begin{equation}
    \nu^\epsilon_{th,pe} \propto n_e\, T_e^{-3/2},
    \label{eq:eps_pe}
\end{equation}
so that at $T_e = 18$~MK and $n_e = 10^{16}$~m$^{-3}$, characteristic of flare plasma at the formation heights of the hot Fe lines, the proton--electron equilibration timescale is $\tau^\epsilon_{pe} = 1/\nu^\epsilon_{th,pe} \approx 100$~s~\citep{Russell2025ApJ...990L..39R}.
The temperature-isotropisation rates of the two species are related to this equilibration rate by
\begin{align}
    \nu_T^p &\approx 6\, \nu^\epsilon_{th,pe}, \label{eq:nuTp}\\
    \nu_T^e &\approx (m_p/m_e)^{1/2}\, \nu_T^p, \label{eq:nuTe}
\end{align}
where the coefficients are those appropriate to the null hypothesis~\citep{NRL}. At the same plasma conditions, proton temperature anisotropies relax on $\tau_T^p = 1/\nu_T^p \approx 15$--$20$~s, whereas the electron temperature isotropises on $\tau_T^e = \tau_T^p / \sqrt{m_p/m_e} \approx 0.4$~s. The strong asymmetry between the two, a factor of $\sqrt{m_p/m_e} \approx 43$, justifies treating the electron distribution as isotropic (a single $T_e$) while allowing distinct parallel and perpendicular ion temperatures, $T_{i\parallel} \neq T_{i\perp}$, to develop.
In a thermal plasma, minor ion species such as the Fe ions observed by EIS are strongly coupled to the protons by Coulomb collisions, with the relaxation of their distribution functions toward the proton distribution occurring on timescales short compared to both $\tau^\epsilon_{pe}$ and $\tau_T^p$. We therefore treat the proton temperature as a proxy for the temperatures of the emitting Fe ions in all quantitative estimates of ion-temperature anisotropy.

\subsubsection*{CGL double-adiabatic prediction for expanding flux tubes}
The comparison of ion-heating mechanisms in the Discussion uses the Chew--Goldberger--Low (CGL) double-adiabatic model~\citep{Chew1956}, which applies when collisions and heat fluxes are neglected. For a magnetised plasma parcel advected along a flux tube, the two CGL invariants combine to yield
\begin{equation}
    \frac{T_\parallel}{T_\perp} \propto \frac{n^2}{B}\,,
    \label{eq:CGL}
\end{equation}
where $n$ is the plasma density and $B$ the magnetic field strength along the flux tube. For chromospheric evaporation into a post-flare flux tube that expands with height, $B$ decreases along the trajectory of the upflowing plasma. Eq.~(\ref{eq:CGL}) therefore predicts $T_\parallel/T_\perp$ to increase during the evaporation, naturally developing $T_{\parallel} > T_{\perp}$, provided the density of the upflowing parcel does not decrease faster than $B^{1/2}$.

\begin{table*}
\caption{Anisotropic broadening parameters derived from the bi-Maxwellian model fit before the GOES soft X-ray peak. Values are quoted as bootstrap medians with 95\% confidence intervals.}
\label{tab:anisotropy_fit_all_before}
\centering

\scriptsize
\setlength{\tabcolsep}{4pt}
\renewcommand{\arraystretch}{1.15}

\begin{tabular}{@{}l c c c c c c@{}}
\toprule
\textbf{Ion} & $T_f$ (MK) & $N$ & $\mu$ & $p$ & $T_{\perp}$ (MK) & $T_{\parallel}$ (MK) \\
\midrule

\multicolumn{7}{@{}l}{\textbf{C1 flares}}\\
\midrule
\ion{Fe}{X} 184.54 & 0.9 &
434 & \celltwolines{-0.223}{-0.947,\,0.483} & 0.791 &
\celltwolines{2.75}{0.49,\,5.19} &
\celltwolines{3.55}{2.74,\,11.60} \\

\ion{Fe}{XII} 192.39 & 1.3 &
640 & \celltwolines{-0.088}{-0.666,\,0.195} & 0.839 &
\celltwolines{2.82}{1.61,\,3.41} &
\celltwolines{3.09}{2.58,\,5.60} \\

\ion{Fe}{XIV} 264.79 & 1.9 &
657 & \celltwolines{-0.136}{-0.635,\,0.170} & 0.029 &
\celltwolines{3.35}{2.10,\,4.24} &
\celltwolines{3.87}{3.23,\,6.32} \\

\ion{Fe}{XVI} 262.98 & 2.6 &
638 & \celltwolines{-0.096}{-0.527,\,0.308} & 0.019 &
\celltwolines{4.71}{3.14,\,6.08} &
\celltwolines{5.21}{4.12,\,7.65} \\

\midrule
\multicolumn{7}{@{}l}{\textbf{C2--3 flares}}\\
\midrule
\ion{Fe}{X} 184.54 & 0.9 &
411 & \celltwolines{0.121}{-0.737,\,2.729} & 0.025 &
\celltwolines{4.44}{1.35,\,6.88} &
\celltwolines{3.96}{1.43,\,7.73} \\

\ion{Fe}{XII} 192.39 & 1.3 &
518 & \celltwolines{0.014}{-0.551,\,0.839} & 0.054 &
\celltwolines{2.99}{1.79,\,4.24} &
\celltwolines{2.95}{1.96,\,4.76} \\

\ion{Fe}{XIV} 264.79 & 1.9 &
548 & \celltwolines{-0.104}{-0.625,\,0.138} & 0.731 &
\celltwolines{3.72}{2.27,\,4.56} &
\celltwolines{4.15}{3.54,\,6.91} \\

\ion{Fe}{XVI} 262.98 & 2.6 &
554 & \celltwolines{-0.095}{-0.582,\,0.359} & 0.782 &
\celltwolines{5.00}{3.06,\,5.79} &
\celltwolines{5.53}{3.95,\,8.39} \\

\midrule
\multicolumn{7}{@{}l}{\textbf{C$>$4 flares}}\\
\midrule
\ion{Fe}{X} 184.54 & 0.9 &
316 & \celltwolines{-0.722}{-0.893,\,0.050} & 0.001 &
\celltwolines{2.09}{0.99,\,4.10} &
\celltwolines{7.50}{2.77,\,10.90} \\

\ion{Fe}{XII} 192.39 & 1.3 &
434 & \celltwolines{-0.361}{-0.676,\,0.145} & 0.002 &
\celltwolines{2.59}{1.80,\,3.43} &
\celltwolines{4.05}{2.74,\,6.10} \\

\ion{Fe}{XIV} 264.79 & 1.9 &
422 & \celltwolines{-0.489}{-0.638,\,0.239} & 0.001 &
\celltwolines{3.00}{2.44,\,4.09} &
\celltwolines{5.86}{2.93,\,7.22} \\

\ion{Fe}{XVI} 262.98 & 2.6 &
420 & \celltwolines{-0.370}{-0.583,\,0.352} & 0.001 &
\celltwolines{4.53}{3.54,\,5.80} &
\celltwolines{7.19}{4.06,\,9.39} \\

\ion{Fe}{XXIII} 263.77 & 14.1 &
79 & \celltwolines{0.027}{-0.471,\,0.501} & 0.262 &
\celltwolines{29.33}{16.38,\,39.02} &
\celltwolines{28.55}{22.23,\,35.58} \\

\ion{Fe}{XXIV} 255.10 & 17.8 &
99 & \celltwolines{-0.137}{-0.419,\,0.386} & N/A &
\celltwolines{24.29}{15.86,\,34.08} &
\celltwolines{28.16}{22.27,\,31.32} \\

\bottomrule
\end{tabular}
\end{table*}

\begin{table*}
\caption{Anisotropic broadening parameters derived from the bi-Maxwellian model fit after the GOES soft X-ray peak. Values are quoted as bootstrap medians with 95\% confidence intervals.}
\label{tab:anisotropy_fit_all_after}
\centering

\scriptsize
\setlength{\tabcolsep}{4pt}
\renewcommand{\arraystretch}{1.15}

\begin{tabular}{@{}l c c c c c c@{}}
\toprule
\textbf{Ion} & $T_f$ (MK) & $N$ & $\mu$ & $p$ & $T_{\perp}$ (MK) & $T_{\parallel}$ (MK) \\
\midrule

\multicolumn{7}{@{}l}{\textbf{C1 flares}}\\
\midrule
\ion{Fe}{X} 184.54 & 0.9 &
411 & \celltwolines{-0.258}{-0.847,\,0.774} & 0.280 &
\celltwolines{2.47}{0.95,\,4.86} &
\celltwolines{3.34}{1.94,\,7.47} \\

\ion{Fe}{XII} 192.39 & 1.3 &
637 & \celltwolines{-0.056}{-0.628,\,0.135} & 0.323 &
\celltwolines{2.83}{1.76,\,3.24} &
\celltwolines{3.00}{2.56,\,5.28} \\

\ion{Fe}{XIV} 264.79 & 1.9 &
632 & \celltwolines{-0.087}{-0.499,\,0.096} & 0.208 &
\celltwolines{3.35}{2.42,\,3.96} &
\celltwolines{3.67}{3.27,\,5.36} \\

\ion{Fe}{XVI} 262.98 & 2.6 &
640 & \celltwolines{-0.011}{-0.422,\,0.276} & 0.215 &
\celltwolines{4.88}{3.32,\,5.58} &
\celltwolines{4.93}{3.92,\,6.50} \\

\midrule
\multicolumn{7}{@{}l}{\textbf{C2--3 flares}}\\
\midrule
\ion{Fe}{X} 184.54 & 0.9 &
356 & \celltwolines{0.180}{-0.462,\,4.928} & 0.051 &
\celltwolines{3.66}{1.64,\,8.15} &
\celltwolines{3.10}{1.17,\,4.45} \\

\ion{Fe}{XII} 192.39 & 1.3 &
470 & \celltwolines{-0.098}{-0.531,\,0.578} & 0.118 &
\celltwolines{2.89}{1.83,\,3.51} &
\celltwolines{3.20}{2.01,\,4.26} \\

\ion{Fe}{XIV} 264.79 & 1.9 &
508 & \celltwolines{-0.122}{-0.473,\,0.004} & 0.041 &
\celltwolines{3.60}{2.40,\,3.94} &
\celltwolines{4.10}{3.61,\,5.04} \\

\ion{Fe}{XVI} 262.98 & 2.6 &
513 & \celltwolines{-0.157}{-0.542,\,1.104} & 0.032 &
\celltwolines{4.86}{3.21,\,7.38} &
\celltwolines{5.77}{3.32,\,7.43} \\

\midrule
\multicolumn{7}{@{}l}{\textbf{C$>$4 flares}}\\
\midrule
\ion{Fe}{X} 184.54 & 0.9 &
351 & \celltwolines{-0.455}{-0.873,\,0.229} & 0.003 &
\celltwolines{3.15}{1.24,\,4.55} &
\celltwolines{5.78}{2.79,\,11.84} \\

\ion{Fe}{XII} 192.39 & 1.3 &
484 & \celltwolines{-0.333}{-0.691,\,0.426} & 0.001 &
\celltwolines{2.72}{1.98,\,3.63} &
\celltwolines{4.07}{2.18,\,7.27} \\

\ion{Fe}{XIV} 264.79 & 1.9 &
487 & \celltwolines{-0.347}{-0.629,\,0.478} & 0.001 &
\celltwolines{3.20}{2.44,\,4.51} &
\celltwolines{4.91}{2.67,\,7.42} \\

\ion{Fe}{XVI} 262.98 & 2.6 &
486 & \celltwolines{-0.299}{-0.494,\,0.189} & 0.001 &
\celltwolines{4.53}{3.87,\,5.35} &
\celltwolines{6.46}{4.08,\,7.93} \\

\ion{Fe}{XXIII} 263.77 & 14.1 &
125 & \celltwolines{-0.163}{-0.498,\,0.533} & 0.030 &
\celltwolines{20.25}{13.66,\,29.33} &
\celltwolines{24.21}{17.55,\,30.11} \\

\ion{Fe}{XXIV} 255.10 & 17.8 &
134 & \celltwolines{-0.237}{-0.673,\,0.822} & 0.070 &
\celltwolines{19.06}{9.40,\,29.49} &
\celltwolines{24.98}{15.11,\,31.70} \\

\bottomrule
\end{tabular}
\end{table*}

\subsection{Bi-Maxwellian fit parameters for all flare classes}
\label{app:fitting}

Tables~\ref{tab:anisotropy_fit_all_before}--\ref{tab:anisotropy_fit_all_after} list the anisotropic broadening parameters derived from the bi-Maxwellian fits for each flare class separately. Values are quoted as bootstrap medians with 95\% confidence intervals.





\end{appendices}


\bibliography{sn-bibliography}

\end{document}